\documentstyle[twoside,fleqn,my_espcrc2,epsf]{article}

\newcommand{\ewxy}[2]{\setlength{\epsfxsize}{#2}\epsfbox[10 60 640 570]{#1}}

\newcommand{\be}{\begin{equation}}
\newcommand{\ee}{\end{equation}}
\newcommand{\order}{{\cal O}}

\newcommand{\fig}[1]{Fig.~\ref{#1}}

\newcommand{\nl}{\nonumber \\}

\newcommand{\mv}{{M_V}}
\newcommand{\mps}{{M_{PS}}}
\newcommand{\mpsq}{{M_{PS}^2}}

\newcommand{\mssbsq}{{M_{s\bar s}^2}}
\newcommand{\cpt}{{\chi PT}}

\newcommand{\ie}{{\rm i.e.\ }}
\newcommand{\mev}{{\rm MeV}}
\newcommand{\gev}{{\rm GeV}}

\newcommand{\AmS}{{\protect\the\textfont2
  A\kern-.1667em\lower.5ex\hbox{M}\kern-.125emS}}

\title{A comparison of Clover and Wilson spectroscopy in the presence of
dynamical quarks}

\author{S. Collins, R.G. Edwards, U.M. Heller, and 
J. Sloan\address{SCRI, Florida State University}
}

\begin{document}

\begin{abstract}
We present preliminary results of light hadron spectroscopy using valence, 
tadpole-improved, Clover fermions on an ensemble of gauge configurations 
generated with 2 flavors of staggered fermions at $\beta=5.6$.  We compare 
the slope and intercept of the curve $\mv$ vs\@. $\mpsq$ for Clover and
Wilson fermions.  
We show that a higher order
chiral perturbation theory ansatz works very well for chiral extrapolations.
\end{abstract}

\maketitle

\section{INTRODUCTION}
In this talk, we present preliminary results from a detailed 
comparison of Clover
and Wilson light spectroscopy at a moderate inverse lattice spacing 
(about $2~\gev$).
%
We discuss two analytic methods; 1) (tadpole-improved)
Symanzik improvement of the fermion action and 2) higher order chiral 
extrapolations.  In addition, we use correlated fits to multiple smearing
functions and an automated plateau finder with different selection criteria
in an attempt to reduce systematic fitting errors.  

\section{THE SIMULATION}
The gauge configurations 
were generated by the 
HEMCGC collaboration~\cite{hemcgc-ens}, with $\beta=5.6$ and $2$ 
flavors of staggered fermions at a mass of $am=.01$.
Inversions were performed on $100$ configurations separated by 20 trajectories,
fixed to Coulomb gauge.
We use tadpole improvement in our calculations; $u_0 = .867$ is 
obtained from the plaquette.

Quark propagators were inverted for each $\kappa$ using a gaussian source; at
some $\kappa$'s a wall source was also used.  
Mixed $\kappa$ hadrons were formed between propagators generated in the same
run.
For the V and PS mesons, each spatial (mesonic) smearing
yields two meson interpolating fields; one with spin structure $\Gamma$ 
and the other with spin structure  $\Gamma*\gamma_0$.  Similarly, we measure
two different $\Delta^+$ states; $s_z=3/2$ and $s_z=1/2$.  

\section{FITTING}
A technique we find very useful is to perform a correlated multi-state fit 
to multiple correlation functions which have the same quantum numbers.  Two 
types of fit are to a ``vector'' of smeared source and local sink and 
to a ``matrix'' of smeared source and sink correlations.  Even if only
a single energy is used in the fit ansatz, multiple smearings can
help eliminate the problem of spurious plateaus because the multiple 
propagators will have different overlaps with excited states.
A notation we find useful is to call a single state fit involving two smearings 
a ``1c2s fit.''

In choosing our fits, we have tried to avoid human intervention by using
automated plateau finders.  I discuss two of our criteria here;
"C", which tends to pick very short plateaus (conservative) and "D",
which prefers long plateaus (aggressive).  Because correlation functions
are noisier at later times, the trade-off between different
criteria is between statistical uncertainties and systematic fitting 
errors; an aggressive criterion will give smaller error bars but may
have more excited state contamination in the ground state mass.  The goal 
is to use the most aggressive
criterion for which the statistical uncertainty dominates fitting errors.
The definition of C is min($T_{min} + \chi^2/N_{prop}$); for D it is
max($Q*N_{dof}/\delta M$), where $N_{prop}$ is the number of 
propagators being
fit, Q is the confidence level of the fit, and $\delta M$ is the fit 
uncertainty in the mass.


To perform chiral fits, we generated 201 bootstrap ensembles of gauge 
configurations, with corresponding sets of propagators.
After choosing a plateau for a particular fit, we repeated the fit on 
each bootstrap ensemble, resulting in an ensemble of 201 mass and amplitude
values.  
Note that, unlike \cite{gf11}, the plateau was kept fixed over
all bootstrap ensembles.  We would like to implement the method of \cite{gf11},
which allows the plateau to vary between bootstrap ensembles,
as this will move systematic fitting errors into bootstrap fluctuations where
they can be statistically quantified.  

Note that the mass ensembles depend both on the type of fit (\ie 
1c1s vs\@. 1c2s)
and on the fit criterion used (\ie C vs\@. D).  A too aggressive criterion
results in large $\chi^2$ in the chiral fit; this is because excited
state effects (systematic errors) are about the same size as the 
statistical errors. This allows investigation of systematic fitting errors;
our results should be consistent between different criteria and types of fits.

\section{$\kappa_c$}
\begin{figure}[htb]
\centerline{\ewxy{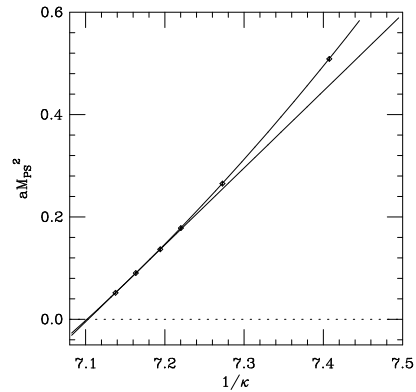}{60mm}}
\caption{
Linear and quadratic fits of $\mpsq$ vs\@. $1/\kappa_c$ for clover action.
}
\label{fig:kcrit}
\vskip -5mm
\end{figure}

Our first example of a higher order chiral extrapolation is in
the determination of $\kappa_c$.  In \fig{fig:kcrit}, we show $\mpsq$ 
vs\@. $1/\kappa$ for the clover action, along with quadratic and linear fits.  
The quadratic fit includes all six points shown, while a linear fit is possible
only for the first three points.  The values of $\kappa_c$ extracted from these
fits are .140809(16) and .140782(16), respectively.  A typical problem when
determining $\kappa_c$ is that it ``runs away''; linear extrapolations to
$\kappa_c$ change as lighter $\kappa$'s are included.  This does not seem
to be the case for the quadratic fit; omitting the lightest two $\kappa$'s
from the fit results in a $\kappa_c$ of .14079(2), which is consistent with
the full quadratic fit.  Because of this we trust the quadratic fit; even our
lightest $\kappa$ values are probably too heavy to be in the linear regime.

\section{MESONS}
\begin{figure}[htb]
\centerline{\ewxy{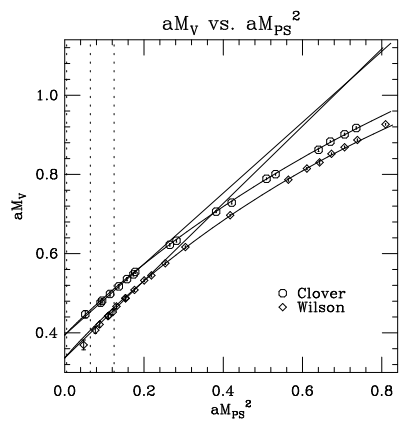}{60mm}}
\caption{
Quadratic and cubic chiral fits to $\mv$ for clover and Wilson actions.
}
\label{fig:rho_both}
\vskip -5mm
\end{figure}

In \fig{fig:rho_both} we present our results
for $\mv$ vs\@. $\mpsq$ for the Wilson and clover actions; both
degenerate and non-degenerate $\kappa$ combinations are included.  
The vertical dotted
lines represent the $\rho$, $K^*$ and $\phi$ masses.  
The $\rho$ and $K^*$
were chosen by requiring that the ratio $\mv/\mps$ match experiment.  The
pseudoscalar mass corresponding to the $\phi$ was chosen 
using the 
formula $\mssbsq = 2m_{K^*}^2 - m_\pi^2$.
The mass values for both V and PS were obtained from 1c2s vector fits using
criterion D; the two smearings were the same spatially and differed by 
$\gamma_0$ in spin structure.

%

Chiral fits were performed using four different ans\"atze; 
\begin{eqnarray*}
{\rm 2-1} &:& C_0 + C_2\mps^2                                             \nl
{\rm 3-1} &:& C_0 + C_2\mps^2 + C_3\mps^3                                 \nl
{\rm 4-2} &:& C_0 + C_2\mps^2 \hphantom{ + C_3\mps^3 }
                                          + C_4\mps^4                     \nl
{\rm 4-1} &:& C_0 + C_2\mps^2 + C_3\mps^3 + C_4\mps^4
\end{eqnarray*}
where the first number indicates the order of the 
polynomial and the second is the number
of coefficients set to zero.  The 2-1 (quadratic) ansatz is just lowest 
order $\cpt$; the 4-2 (quartic) ansatz introduces a higher order term
that goes like $m_q^2$.  The cubic term in the 3-1 (cubic) and 4-1 (full
quartic) ans\"atze 
were motivated by the one loop $\cpt$ correction to baryon masses; at 
this conference we learned of \cite{chiral_vector}, in which one loop 
$\mps^3$ corrections to the vector meson mass are derived.  

The results of chiral fits were qualitatively similar for Wilson and clover.
In both cases, a quadratic fit breaks down after the sixth point is
included, while both the cubic and quartic (4-2) fits were able to 
include many more points (typically 11-18 points).  The cubic fit,
however, seemed to do a better job of describing data at even higher
masses which could not be fit; it and the quadratic fits are shown
in \fig{fig:rho_both}.
The full quartic (4-1) ansatz only allowed
one or two more points to be fit than either a cubic or quartic ansatz; this
is a sign that there is no signal to resolve the additional
parameter.  The numerical results of our fits are presented in 
Tab.~\ref{tab:rho_fits}.  We have also performed fits
using only degenerate $\kappa$ combinations and obtain consistent results.
Note that we can resolve the difference between $M_\rho$ at $M_\pi=.18M_\rho$
and at $M_\pi = 0$; we obtain $aM_\rho = .341(4), .400(4)$ for Wilson and
clover, respectively.

\begin{table}[thb]
%
\caption{Chiral fit results for $\mv$ vs\@. $\mps$. `*' indicates
light $\kappa$'s were omitted.}
\label{tab:rho_fits}
\begin{tabular}{c|cc|cc}
\hline
Fit &\multicolumn{2}{|c|}{Wilson} & \multicolumn{2}{|c}{Clover} \\
\cline{2-5}
    & $C_0$ & $C_2$ & $C_0$ & $C_2$\\
\hline
2-1  & .337(6)  &   -     & .395(6)  &   -     \\
3-1  & .336(4)  & 1.23(2) & .394(4)  & 1.10(3) \\
4-2  & .337(6)  & 1.13(5) & .403(6)  & 0.93(2) \\
3-1* & .334(4)  & 1.25(2) & .393(4)  & 1.10(3) \\
\hline
\end{tabular}
\end{table}

Note that the cubic fits are consistent with and have 
smaller error bars than the quadratic fits.  This indicates that the 
higher mass points, which have smaller fluctuations, are controlling the 
fit, \ie the lightest
(and most expensive) $\kappa$'s are too noisy to contribute useful information.
To test this, we omitted all $\kappa$ combinations lighter than the third
degenerate $\kappa$ (6 points for Wilson and 4 for clover) and repeated the
3-1 and 2-1 fits.  The quadratic ansatz only fit one degree of freedom (3 
points) for clover and none for Wilson.  The cubic fits, however, succeeded
to the same $\kappa_{max}$ and gave results with the same error 
bars. {\it This indicates that it was unnecessary to 
simulate below the strange quark mass in order to calculate the 
physical $\rho$ mass.} 

\section{BARYONS}
For baryons, we have not yet analyzed our Wilson results.  In addition, we
have results with wall smearings at only two of our six lightest $\kappa$'s.
We will correct these deficiencies in the near future.


\begin{figure}[htb]
\centerline{\ewxy{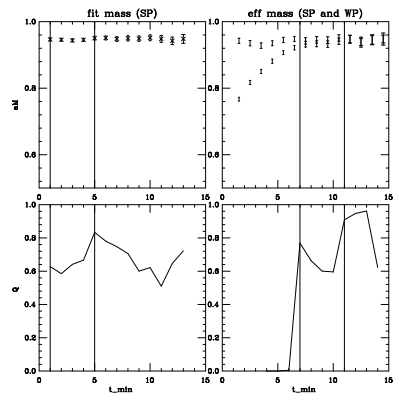}{60mm}}
\caption{
Comparison between fit criteria and between single and multi-propagator
fits for the $\kappa=.1375$ nucleon.  Vertical lines indicate $T_{min}$ 
chosen by criteria D and C (D is always leftmost).  Upper right plot is
superposition of effective mass for shell and wall sources; upper left plot
is mass values for 1c1s fit as function of $T_{min}$.
}
\label{fig:smeared_wall}
\vskip -5mm
\end{figure}

\begin{figure}[htb]
\centerline{\ewxy{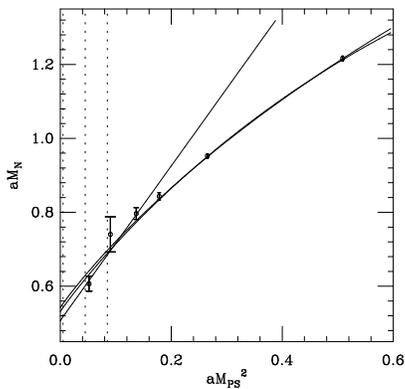}{60mm}}
\caption{
Chiral fits to $M_{nucleon}$ generated from
 1c1s fits and criterion C.
}
\label{fig:proton_good}
\vskip -5mm
\end{figure}

One interesting result for the nucleon was that we were unable to obtain
good chiral fits when criterion D was used for 1c1s propagator fits.
We interpret this as an example of too aggressive a fit 
criterion leading to statistically significant excited state contamination
of the ground state energy.
This can be corrected by either using a
less aggressive fit criterion (\ie C), or by including a second 
smearing (\ie wall).  This can be seen in \fig{fig:smeared_wall}; the
1c1s criterion D fit uses $T_{min}=1$!  This is changed to 5 and 7 for
the C criterion and 1c2s fit, respectively. In \fig{fig:proton_good} we
show that using criterion C results in successful chiral fits, but that
C is probably too conservative.  The $\Delta$
behaves similarly; criterion D gives mass values which can
not be fit, while criterion C works well.  As with the vector, a
cubic ansatz allows much heavier quark masses to be included.

\section{CONCLUSIONS}
Our most striking result is that a cubic ansatz for chiral extrapolations
of the vector mesons works so well, and over such a large mass range.  Our
results seem to indicate that the two $\kappa$ values lighter 
than $\kappa_{strange}$ were unnecessary, since their omission from fits 
caused negligible changes.  This is especially important given that the light
$\kappa$'s are both the most expensive and the most sensitive to finite 
volume effects.  A potential drawback to this approach was our inability to
clearly distinguish between cubic and quartic ans\"atze, and the lack of 
a strong signal to justify including both terms at the same time.

\begin{table}[thb]
%
\caption{Comparison of dimensionless ratios}
\label{tab:expt}
\begin{tabular}{cccc}
\hline
Observable & Wilson & Clover & Expt. \\
\hline
$J_d$                             & .425(6)  &  .444(5)  & .499 \\
($M_\Omega-M_\Delta)/M_{K^*}$     &    -     &  .39(5)   & .491 \\
$M_\Omega/M_{K^*}$                &    -     &  .575(11) & .536 \\
\hline
\end{tabular}
\end{table}

Our comparison between Wilson and clover actions has only been completed
for mesons; the results were reasonable.  For both the intercept and slope
of $\mv$ vs\@. $\mpsq$, we saw $10$-$20\%$ discrepancy.  This is consistent
with a naive estimate of $500~\mev * a \approx .25$ for Wilson 
discretization errors, and suggests that the \order($a^2$) clover 
discretization errors are roughly $5\%$ at this lattice spacing.  A comparison
between clover results and experiment is less encouraging; as
can be seen from Tab.~\ref{tab:expt}, the
ratio $J_d = M_{K^*}*(M_{K^*}-M_\rho)/(M_K^2-M_\pi^2)$ is 10\% too small and the
ratio $(M_\Omega-M_\Delta)/M_{K^*}$ is 20\% too small.  We hope that this effect
is due to either finite volume or incorrect dynamical fermion content; if this
is not the case then either clover is not correctly removing \order(a) errors 
or there is some source of error which we do not understand.

\section*{ACKNOWLEDGEMENTS}
We thank Rajan Gupta, Steve Sharpe, and Maarten Golterman for useful
conversations.
The computer simulations were performed on the CM-2 at SCRI.
This research was supported by DOE contracts DE-FG05-85ER250000
and DE-FG05-92ER40742.


\begin{thebibliography}{9}
\bibitem{hemcgc-ens} K.~Bitar et al.,
   Nucl. Phys. B (Proc. Suppl.) 26 (1992) 259; {Phys. Rev.} D46 (1992) 2169.
\bibitem{gf11} F. Butler et al.,
   Nucl. Phys. B421 (1994) 217; B430 (1994) 179.
\bibitem{chiral_vector}   E.~Jenkins et al., 
   UCSD-PTH-95-08 (1995), hep-ph 9506356
\end{thebibliography}
\end{document}